\documentclass [prl,nofootinbib]  {revtex4}
\usepackage{amssymb}

\def\no{\, : \,}

\newcommand{\be}{\begin{equation}}
\newcommand{\ee}{\end{equation}}
\newcommand{\bea}{\begin{eqnarray}}
\newcommand{\eea}{\end{eqnarray}}

\begin{document}

\title{%
From noncommutative $\kappa$--Minkowski to Minkowski space-time}
\author{Laurent Freidel}
\affiliation{Perimeter
Institute, 31 Caroline Street North Waterloo, N2L
2Y5, Ontario, Canada.}
\author{Jerzy Kowalski-Glikman}
\affiliation{Institute for Theoretical Physics, University of Wroc\l{}aw, Pl.\ Maxa Borna 9, Pl--50-204 Wroc\l{}aw, Poland.}
\author{Sebastian Nowak}
\affiliation{Institute for Theoretical Physics, University of Wroc\l{}aw, Pl.\ Maxa Borna 9, Pl--50-204 Wroc\l{}aw, Poland.}

\begin{abstract}
We show that free $\kappa$-Minkowski space field theory is equivalent to a relativistically invariant, non local, free field theory on Minkowski space-time.
 The field theory we obtain has in spectrum a
relativistic mode of arbitrary mass $m$ and a Planck mass tachyon.
We show that while the energy momentum for the relativistic mode is
essentially the standard one, it diverges for the tachyon, so that
there are no asymptotic tachyonic states in the theory. It also
follows that the dispersion relation is not modified, so that, in
particular, in this theory the speed of light is energy-independent.
\end{abstract}

\maketitle

The $\kappa$-Minkowski space \cite{kappaM1}, \cite{kappaM2} is a
noncommutative space-time, in which time and space coordinates
satisfy the following Lie-type commutation
relation \begin{equation}\label{1}
 [\hat x_0, \hat x_i] = - \frac{i}{\kappa}\, \hat x_i.
\end{equation}
As shown in \cite{kappaM1} this spacetime arose quite naturally from
$\kappa$-Poincar\'e algebra \cite{qP}. More recently it played a
central role in Doubly Special Relativity research program
\cite{Amelino-Camelia:2000ge}, \cite{Amelino-Camelia:2000mn},
\cite{rbgacjkg} (for recent review see
\cite{Kowalski-Glikman:2004qa}.)

To investigate properties of $\kappa$-Minkowski spacetime it seemed
quite natural to construct field theories on it and then discuss
their physical properties. This endeavor was undertaken by many
authors \cite{Kosinski:1999ix}, \cite{Kosinski:1998kw},
\cite{Amelino-Camelia:2001fd}, \cite{Kosinski:2003xx},
\cite{Daszkiewicz:2004xy} (see also \cite{Dimitrijevic:2004nv}).
Unfortunately, in spite of this effort our present understanding of
physics (on the background) of $\kappa$-Minkowski spacetime is still
rather incomplete.

The aim of this note is to fill this gap. Using star product for
$\kappa$-Minkowski space whose construction we will briefly describe
below, we will be able to recast free scalar field theory on this
non-commutative spacetime to the form of surprisingly simple, {\em
equivalent} scalar field theory living on ordinary Minkowski space.
The action of this theory contains infinite powers of the wave
operator, reflecting non-locality of the original theory, and
reminding somehow the situation discussed in context of p-adic
string theory (see \cite{Moeller:2002vx} and references therein).

In this letter we will concentrate on deriving the relevant field
theory on Minkowski space skipping many technical points, especially
the relation to the equivalent construction on $\kappa$-Minkowski
space, leaving them to a longer paper \cite{LJStoappera}.

Let us start with a simple, but crucial observation. The algebra
(\ref{1}), in the context of DSR understood as an (sub)algebra of
tangent vectors to de Sitter space of momenta
\cite{Kowalski-Glikman:2003we} can be regarded a Lie algebra
$b_\kappa$ of a Lie group $B_\kappa$ that we will call below Borel
algebra and group, respectively. This group arises naturally in
Iwasawa decomposition of the $SO(4,1)$ into ``Lorentz'' and
``translational'' parts, and can be therefore identified with (a
portion of) de Sitter space of momenta. Thus the ordered plane waves
on $\kappa$-Minkowski spacetime \cite{Amelino-Camelia:1999pm}
\begin{equation}\label{1a}
\hat{e}_k \equiv \no e^{i k_\mu \hat{x}^\mu}\no =  e^{i \mathbf{k}\,
\hat \mathbf{x}}\, e^{-ik_0 \hat x_0}.
\end{equation}
can be equivalently regarded as Borel group elements. Accordingly,
the composition of plane waves, usually understood in terms of the
coproduct of $\kappa$-Poincar\'e algebra, can be regarded as a
simple group elements product. Explicitely
\be
\hat{e}_k \hat{e}_p = \hat{e}_{kp}, \quad  kp\equiv (k_0 + p_0, k_i +
e^{-\frac{k_0}{\kappa}}\, p_i)
\ee
Note that Borel group can be
coordinatized by labels $k$ in the plane waves. These coordinates
correspond to the ``cosmological coordinates'' on (the portion of)
de Sitter space of momenta \cite{Kowalski-Glikman:2003we}.

We now analyze the property of Fourier transformation on
$\kappa$-Minkowski space. All the results obtained below are natural
extension of the ones obtained in \cite{Freidel:2005ec},
\cite{Freidel:2005bb} in the context of 3-dimensional
$\kappa$-Minkowski space.

First in order to do physics it is convenient to map, under a Weyl
map,  the  algebra of fields on non commutative $\kappa$-Minkowski
space into the space of usual fields on Minkowski spacetime equipped
with a star-product. The key point is that among all possible
choices of Weyl maps one is preferred, as explained in more details
in \cite{LJStoappera}. This is the one which preserves the Lorentz
invariant bicovariant calculus or equivalently the one under which
the action of the Poincar\'e group becomes linear.
 This Weyl map is fully characterized by its action on plane waves:
 it maps  the $\kappa$-Minkowski plane waves
$e_{k}$ which form a non abelian group $B_{\kappa}$ to
usual plane waves $e^{iP(k)_{\mu} x^{\mu}}$ as follows
\begin{eqnarray}
P: B_{\kappa} &\to& R^{4} \nonumber\\
k&\mapsto& P(k)\label{2} \end{eqnarray} where $P(k)$
\begin{eqnarray}
 {P_0}(k_0, \mathbf{k}) &=&  \kappa\sinh
{\frac{k_0}{\kappa}} + \frac{\mathbf{k}^2}{2\kappa}\,
e^{  \frac{k_0}{\kappa}} \nonumber\\
 P_i(k_0, \mathbf{k}) &=&   k_i \, e^{  \frac{k_0}{\kappa}}
 \label{3}
\end{eqnarray}
are naturally understood as labelling points in a de Sitter space of
radius $\kappa$ (with $ {P_4}(k_0, \mathbf{k}) =  \kappa\cosh
{\frac{k_0}{\kappa}} - \frac{\mathbf{k}^2}{2\kappa}\,
e^{  \frac{k_0}{\kappa}}$).

 This choice of map is dictated by the bicovariant
differential calculus \cite{5dcalc1}, which is the starting point of
construction of Poincar\'e invariant field theory on
$\kappa$-Minkowski space. Note that not all of $R^{4}$ is in the
image of $P$ since only the points for which $P_{4}^{2}\equiv
\kappa^2 -\mathbf{P}^{2} +P_{0}^{2} \geq0 $ are reached. Moreover
this mapping covers only half of $dS_{4}$, the one for which
$P_{0}+P_{4}>0$. In order to define a notion of Fourier
transformation which is covariant under Lorentz transformation we
need to have a space of momenta which is without boundaries. This is
possible if one interpret the image of $P$ to be the quotient of de
Sitter space by the $Z_{2}$ identification $P_{A}\to -P_{A}$. This
space $dS_{4}/Z_{2}$ is usually called elliptic de Sitter space.
Since the identification is an isometry which  possess no fixed
point in de Sitter the quotient is well defined and is a symmetric
space under the action of $SO(4,1)$. The invariant measure on
$dS_{4}/Z_{2}$ is given by
\begin{equation}
\mathrm{d}P= \frac{1}{2}\mathrm{d}^5P
\delta(P^AP_A-\kappa^2)\label{4}
\end{equation}
A point in $dS_{4}/Z_{2}$ is labeled either by $P_{0}, P_{i}$ or by
$P_{+}, P_{i}$ with the restriction\footnote{This parameterizations
does not cover all of $dS_{4}/Z_{2}$ since we have to exclude the
hyperplane $P_{+}=0$. this hyperplane is exactly the place where the
star product is not well defined. However this hyperplane is of
measure $0$ and this problem should not bother us in the definition
of the Fourier transform.  } $P_{+}\equiv
P_{0}+P_{4}>0$. When written in terms of the group variables the
measure reads\footnote{From now, in order to simplify notations, on we choose the Planck units,
in which the Planck mass scale $\kappa$ as well as the Planck length
scale $1/\kappa$ are equal 1. }

\begin{equation}
 \mathrm{d}P(k) \equiv e^{3k_0}dk_0 d^3\mathbf{k}, \label{5}
 \end{equation}
 which is a left invariant measure on the group
 \begin{equation}
 \quad  \mathrm{d}P(pk)= \mathrm{d}P(k)\label{6}
 \end{equation}
 In order to prove (\ref{5}) let us integrate a function $\phi(P)=\phi(-P)$ on  $dS_{4}/Z_{2}$
 \begin{eqnarray}
 \int dP \phi(P) &=& \int \mathrm{d}^5P \delta(P^AP_A-1)\theta(P_{0}+P_{4}) \frac{1}{2}(\phi(P)+\phi(-P))\nonumber\\
  &=& \int \mathrm{d}P_{+} \mathrm{d}P_{-} \mathrm{d}^{3}\mathbf{P}  \delta(P_{+}P_- - (1-\mathbf{P}^{2})\theta(P_{+})\phi(P)\nonumber\\
  &=& \int \frac{ \mathrm{d}P_{+}(k)}{P_{+}(k)} \, \mathrm{d}^{3}\mathbf{P}(k)   \phi(P(k))\nonumber\\
  &=& \int e^{3k_0}dk_0 d^3\mathbf{k} \, \phi(P(k)),\label{7}
  \end{eqnarray}
  where $P_{\pm} \equiv P_{4}\pm P_{0}$ and $\theta(P)$ is the Heaviside distribution.

Given a field $\phi(x)$ we define its Fourier components
\begin{eqnarray} \tilde{\phi}(P(k))= \int \mathrm{d}^4x
\left(e^{iP(k)\cdot x}\right)^\dagger \star {\phi}(x)\label{8}
\end{eqnarray}
where we introduce a conjugation
\begin{equation}
\left(e^{iP(k)\cdot x}\right)^\dagger\equiv e^{iP(k^{-1})\cdot
x}\label{9}
\end{equation}

This  formula  involves the star product, which is defined to
satisfy
\begin{equation}\label{10}
    e^{iP(k)\cdot x} \star e^{iP(p)\cdot x} \equiv e^{iP(kp)\cdot x}
\end{equation}
where $kp$ on the right hand side is defined by $e_k e_p = e_{kp}$,
i.e. is a label of the group element obtained as a product of group
elements labeled by $k$ and $p$. Similarly we will use the
abbreviation $k^{-1}$ to denote the label of the element $e_{k^{-1}}
= e_k{}^{-1}$. Explicitly, $kp$ has components $(k_0 + p_0, k_i +
e^{-k_0}\, p_i)$, and $k^{-1}$ equals $(-k_0, -k_i \, e^{k_0})$,
from which it follows that  $k^{-1}p$, which is relevant to
the calculation below, is equal $(-k_0+p_0, e^{k_0}(-k_i + p_i))$.

In order to reconstruct  $\phi(x)$ from its Fourier mode we need
to relate this Fourier transform to the usual one where the
inversion formula is known.
 This can be achieved  by using the following key integration identity
which relates the integration of star product of fields to the
integration of usual fields product
\begin{equation} \label{intid}
\int \mathrm{d}^4x \, \phi^\dagger(x) \star \psi(x)
= \int \mathrm{d}^4x
\, \phi^*(x)(1-\partial_4) \psi(x)
\end{equation}
where $*$ denotes
the complex conjugation and
\begin{equation} (1-\partial_4) = \sqrt{
1 +\Box}, \quad \Box= ( - \partial_0^2 +
\partial_i\partial^i)
\end{equation}
is the differential operator arising in the bicovariant differential
calculus \cite{5dcalc1}.

Using this identity the Fourier modes can be written in term of a
usual Fourier transform
\begin{equation}
\tilde{\phi}(P(k))= P_4(k)
\int \mathrm{d}^4x \, e^{-iP(k)\cdot x}  {\phi}(x).
\end{equation}

We can then reconstruct $\phi(x)$ using the usual Fourier
transformation
\begin{equation}
\phi(x)= \int
\frac{\mathrm{d}^4P}{(2\pi)^4} \frac{e^{iP\cdot x}}{P_4(k)}
\tilde{\phi}(P),
\end{equation}
where $\mathrm{d}^4P$
 is the Lebesgue measure.

 This can be written in terms of the right invariant group measure using the fact that
 \begin{equation}
 \mathrm{det}\left(\frac{\partial P_\mu}{\partial k_\nu}\right) = e^{3k_0}P_4(k),
 \end{equation}
 hence
 \begin{equation}
\phi(x)= \int  \frac{\mathrm{d}P(k)}{(2\pi)^4}   e^{iP(k)\cdot x}
\tilde{\phi}(P(k))
\end{equation}

In order to prove (\ref{intid}) it is sufficient to  establish it
for plane waves $\phi(x)= e^{iP(k)\cdot x}, \psi=e^{iP(p)\cdot x}$.
\begin{equation} \int \mathrm{d}^4x \, \left(e^{iP(k)\cdot
x}\right)^\dagger \star e^{iP(p)\cdot x} = \int \mathrm{d}^4x
e^{iP(k^{-1}p)\cdot x}= (2\pi)^4 \delta^4\left(P(k^{-1}p)\right)=
(2\pi)^4
\delta\left(P_0(k^{-1}p)\right)\delta^3\left(\mathbf{P}(k^{-1}p)\right).
\end{equation}
To prove it let us first notice that
\begin{eqnarray}
P_{0}(k^{-1}p) &=&P_{+}(k)P_{0}(p)-P_{0}(k)P_{+}(p) -\frac{P_{i}(k)}{P_{+}(k)}\left(P_{+}(k)P_{i}(p)-P_{i}(k)P_{+}(p)\right),\nonumber\\
P_{i}(k^{-1}p) &=&\frac{P_{+}(k)P_{i}(p)-P_{i}(k)P_{+}(p)}{P_{+}(k)}
\end{eqnarray}
Note that the last term in $P_{0}(k^{-1}p)$ drops out since it is
proportional to $\mathbf{P}(k^{-1}p)$. Then from the relation $
2P_{0}= P_{+}-P_{+}^{-1} + {\mathbf{P}}^{2}/P_{+}$ we compute
\begin{equation}
P_{+}(k)P_{0}(p)-P_{0}(k)P_{+}(p) =
\frac{1}{2P_{+}(p)P_{+}(k)}\left(P_{+}^{2}(k)-P_{+}^{2}(p)+
(P_{+}(k)\mathbf{P}(p))^{2}-(\mathbf{P}(k)P_{+}(p))^{2}\right)
\end{equation}
where again only the first two terms are relevant. It now follows
that
 \begin{eqnarray}
 \delta^4\left(P(k^{-1}p)\right)
&=&\delta\left(
\frac{P_{+}^{2}(k)-P_{+}^{2}(p)}{2P_{+}(p)P_{+}(k)}\right)
\delta^{3}\left(\mathbf{P}(p)-\frac{P_{+}(p)}{P_{+}(k)}\mathbf{P}(k)\right)\\
&=& P_{+}(p)\delta( P_{+}(p) -
P_{+}(k))\delta^{3}(\mathbf{P}(p)-\mathbf{P}(k))
\end{eqnarray}
where we have use the restriction $P_{+}>0$.
And since
$$\left.\frac{\partial P_{0}}{\partial P_{+}}\right|_{P_{i}} =\frac{1}{2}\left( 1+P_{+}^{-2} -
\mathbf{P}^{2}P_{+}^{-2}\right)=\frac{P_{4}}{P_{+}}$$
we can conclude
\begin{equation}
(2\pi)^4 \delta^4\left(P(k^{-1}p)\right) = (2\pi)^4
|P_{4}(p)|\delta( P_{0}(p) -
P_{0}(k))\delta^{3}(\mathbf{P}(p)-\mathbf{P}(k)) = \int
\mathrm{d}^4x \, \left(e^{iP(k)\cdot x}\right)^* |P_{4}(p)|
e^{iP(p)\cdot x}
\end{equation}
which proves the key identity
(\ref{intid}).

The action for a massive field is given by
\begin{eqnarray}
S&=&\int \mathrm{d}^4x\, \frac{1}{2} (\partial_{\mu} \phi)^\dagger \star (\partial^{\mu} \phi)(x) + \frac{m^2}{2} \phi^\dagger\star \phi(x)\label{48}\\
&=&\int \mathrm{d}^4x\, \frac{1}{2} (\partial_{\mu} \phi)^*
(1-\partial_{4}) (\partial^{\mu} \phi)(x) + \frac{m^2}{2}
\phi^*(1-\partial_{4}) \phi(x)\label{49}
\end{eqnarray}
This action is manifestly invariant under Poincar\'e transformations
\begin{equation}
\phi(x) \to \phi(x+a), \quad \phi(x)\to \phi(\Lambda x),
\end{equation}
where $\Lambda$ is a Lorentz transformation.

The formula (\ref{49}) is the main result of our paper. It shows
that with the right choice of star product, that is the one compatible with the bicovariant differential
calculus on $\kappa$-Minkowski space, and careful implementation of
Lorentz symmetry (related to the choice of conjugation (\ref{9})),
the resulting Minkowski space action is simple and covariant. It should be
stressed that the action (\ref{49}) is fully equivalent to the
$\kappa$-Minkowski space action investigated in
\cite{Kosinski:1999ix},\cite{Kosinski:2001ii},
 \cite{Kosinski:2003xx},
\cite{Daszkiewicz:2004xy}, (while our method can be readily applied
in the  case of a bit less natural choice taken in
\cite{Amelino-Camelia:2001fd}, \cite{Dimitrijevic:2004nv} who use slightly different kinetic
term for field theory on $\kappa$-Minkowski space) and therefore it
carries equivalent physical information. Further details of the
construction will be discussed in the forthcoming paper
\cite{LJStoappera}.

Let us investigate the main properties of the action (\ref{49}), in
the case of real field $\phi$. First of all the field equations read
\begin{equation}\label{50}
\sqrt{ 1 +{\Box}}\left(\Box-m^2\right)\, \phi =0
\end{equation}
and describe a mode of mass $m$ along with what seems to be a Planck scale tachyon.

In order to get more insight into physical properties of the theory,
let us compute the energy-momentum tensor. Since the theory has
infinitely many derivatives it is convenient to use the standard
formula for symmetric energy momentum tensor (see
\cite{Moeller:2002vx} and references therein for explanation how to
handle energy momentum tensor in the case of theories with infinite
number of derivatives)
\begin{equation}\label{51}
    T_{\alpha\beta}(y) = \left.\frac2{\sqrt{-g}}\, \frac{\delta S}{g^{\alpha\beta}(y)}\right|_{g=\eta}=\left. \frac{\delta}{g^{\alpha\beta}(y)}\, \int \mathrm{d}^4x\, \sqrt{-g}
    \phi(x)\left( - \Box^g+ m^2\right)\sqrt{1+\Box^g}\,
    \phi(x)\right|_{g=\eta}
\end{equation}
where $\Box^g = \frac1{\sqrt{-g}}
\partial_\mu\sqrt{-g}g^{\mu\nu}\partial_\nu$ is the covariant
wave operator. As it is well know the energy momentum tensor
(\ref{51}) is conserved by construction. Let us consider generic
term that arises in expansion of the right hand side of (\ref{51})
$$
\frac{\delta}{g^{\alpha\beta}(y)}\, \int \mathrm{d}^4x\, \sqrt{-g}
    \phi(x)\, \Box^n\, \phi(x) =
$$
\begin{equation}\label{52}
\frac{\delta}{g^{\alpha\beta}(y)}\, \int
\mathrm{d}^4x\,\phi(x)\,\left(\partial_{\mu_1}\sqrt{-g}g^{\mu_1\nu_1}\partial_{\nu_1}
\ldots\frac1{\sqrt{-g}}\partial_{\mu_n}\sqrt{-g}g^{\mu_n\nu_n}\partial_{\nu_n}\right)\phi(x)
\end{equation}
This expression can be calculated by using the identities
$$
\frac{\delta g^{\mu\nu}}{g^{\alpha\beta}(y)} =
\delta^\mu_\alpha\,\delta^\nu_\beta\,\delta(x-y), \quad \frac{\delta
\sqrt{-g}}{g^{\alpha\beta}(y)} =-\frac12\,
\sqrt{-g}g_{\alpha\beta}(x)\,\delta(x-y), \quad\frac{\delta
}{g^{\alpha\beta}(y)}\,\frac1{\sqrt{-g}} =\frac12\,\frac1{\sqrt{-g}}
\,g_{\alpha\beta}(x)\,\delta(x-y)
$$
Using these expressions and integrating out delta functions we get
the following formula
$$
T^{(n)}_{\alpha\beta}(y) \equiv 2
\left.\frac{\delta}{g^{\alpha\beta}(y)}\, \int \mathrm{d}^4x\,
\sqrt{-g}
    \phi(x)\, (\Box^g)^n\, \phi(x)\right|_{g=\eta} =
    $$
    $$
    =
    \eta_{\alpha\beta}\left(\partial_\mu\phi\partial^\mu\Box^{n-1}\phi
    +\partial_\mu\Box\phi\partial^\mu\Box^{n-2}\phi+\cdots
    +\partial_\mu\Box^{n-1}\phi\partial^\mu\phi\right)(y)
    $$
    $$
    +
    \eta_{\alpha\beta}\left(\Box\phi\Box^{n-1}\phi
    +\Box^2\phi\Box^{n-2}\phi+\cdots
    +\Box^{n-1}\phi\Box\phi\right)
    $$
    $$
    -2\left(\partial_\alpha\phi\partial_\beta\Box^{n-1}\phi
    +\partial_\alpha\Box\phi\partial_\beta\Box^{n-2}\phi+\cdots
    +\partial_\alpha\Box^{n-1}\phi\partial_\beta\phi\right)
    $$
While calculating energy-momentum tensor, in the final formula we
can use equations of motion, i.e., replace $\Box$ with $M^2$, where
$M^2$ equals $m^2$ for particle, and $-1$ for tachyon modes,
respectively. Using this observation we can simplify the formula
above to give
 \begin{equation}\label{53}
T^{(n)}_{\alpha\beta}(y)  =
    \eta_{\alpha\beta}\left(n\,
    M^{2(n-1)}\partial_\mu\phi\partial^\mu\phi + (n-1)\,
    M^{2n}\phi^2\right)-2n\, M^{2(n-1)}\partial_\alpha\phi\partial_\beta\phi.
    \end{equation}
With this formula we are able to calculate energy momentum tensor for
the action of the form
$$
S = \int \mathrm{d}^4x\, \phi f(\Box)\, \phi
$$
 with analytic $f$, to wit
 \begin{equation}\label{54}
T^{(f)}_{\alpha\beta} =
    \eta_{\alpha\beta}\left(f'(M^2)(\partial_\mu\phi\partial^\mu\phi
    +M^{2}\phi^2) - f(M^2)\phi^2\right) -2f'(M^2)\partial_\alpha\phi\partial_\beta\phi.
 \end{equation}
One can easily check that this formula leads to the right answer in
the standard case $S=\frac12\int (-\phi\Box\phi + m^2\phi^2)$. In
our case, (\ref{49}), we have to do with two expressions: $f =
-\frac12\, \Box\,\sqrt{1+\Box}$ for the kinetic and $f =\frac12\,
m^2\,\sqrt{1+\Box}$ for the mass term.

In the particle of mass $m$ case the energy-momentum tensor acquires
just the multiplicative term and reads
\begin{equation}\label{55}
   T_{\alpha\beta}=\sqrt{1+m^2}\left(\partial_\alpha\phi\partial_\beta\phi-\frac12\,\eta_{\alpha\beta}
(\partial_\mu\phi\partial^\mu\phi
    +m^{2}\phi^2)\right)
\end{equation}
note that since $m^2 \ll 1$ the factor in front of expression above
is negligible. Thus the energy and momentum of the free particle
modes is essentially just classical.

In the case of the tachyonic mode, as a result of the presence of
the denominator $\sqrt{1+M^2}$ with $M^2=-1$, the components of
energy-momentum tensor are divergent. This is quite a desirable
feature of the model, since it means that one needs infinite energy
to create a tachyon mode in the free theory. Moreover it is in agreement with the fact that
we initially restricted the momentum space to be such that $P^{2}<\kappa^{2}$ so that the tachyon
is not in the spectra.

In conclusion, we managed to formulate a theory of free scalar field
on Minkowski spacetime, which is  equivalent to a free theory formulated on $\kappa$-Minkowski spacetime. The spectrum
of this theory contains a standard particle of arbitrary mass $m$
along with the Planck mass tachyon, which however requires infinite
energy to be produced.

We don't expect  expect that this latter feature change in the interacting theory since
the `tachyonic' factor $\sqrt{1+\Box}$ comes from the conversion from star product to usual product.
This means that  this term will also factorise in front of the interacting equation of motion and do not propagate.
In other terms this means that we expect this term to appear as a modification of the integration measure over loop momenta but not
as a modification of the propagator in agreement with the analysis performed in \cite{Freidel:2005bb}.

   Of course it is a main challenge to investigate in detail such interacting theory
at the quantum level. In principle the construction of
polynomial interactions does not pose great problem, since we know
from the construction presented here how to use star product and
conjugation to produce Lorentz invariant terms of any order.
For a real field $\phi^{\dagger}= \phi$, and  since
$(\phi \star \phi)^{\dagger}= \phi^{\dagger} \star \phi^{\dagger}$, local non derivative interactions are given by
star product powers of the field like $\phi \star \phi \star \phi \star \phi$ \cite{Amelino-Camelia:2001fd}, \cite{Daszkiewicz:2004xy}.

It should be also stressed that the construction of the energy
momentum tensor (\ref{55}) leads to a standard dispersion relation,
so that, at least in free theory there is no room for energy
dependent speed of light. 
This conclusion contrast with the result
of  \cite{Agostini:2006nc} and can be trace back to their use of a
non covariant differential calculus but  is in agreement with the
discussion of \cite{Kowalski-Glikman:2004qa} on DSR
theory based on $\kappa$-Minkowski space.
 Although it is not excluded that this will change in quantum
interactive theory the effect, if any,  is expected to be extremely small.

{\bf acknowledgment}

For JK-G and SN this work  is partially supported by the grant KBN 1 P03B01828.

\end{document}